\documentclass[aps,prl,twocolumn,showpacs]{revtex4-1} 

\usepackage[dvips]{graphicx}
\usepackage{color}
\usepackage{psfrag}

\usepackage{amsthm}
\usepackage{amsmath}
\usepackage{amsfonts}
\usepackage{amssymb}

\DeclareMathOperator{\arth}{arth}
\renewcommand\Re{\operatorname{Re}}
\renewcommand\Im{\operatorname{Im}}

\renewcommand{\vec}[1]{\mathbf{#1}}

\def\vc{\vec{c}}
\def\vd{\vec{d}}
\def\hvd{\hat{\vec{d}}}
\def\vk{\vec{k}}
\def\vsigma{\boldsymbol{\sigma}}
\def\vdelta{\boldsymbol{\delta}}

\begin{document}
\title{Topological classification of dynamical phase transitions}
\author{Szabolcs Vajna}
\affiliation{Department of Physics and BME-MTA Exotic  Quantum  Phases Research Group, Budapest University of Technology and
  Economics, 1521 Budapest, Hungary}
\author{Bal\'azs D\'ora}
\affiliation{Department of Physics and BME-MTA Exotic  Quantum  Phases Research Group, Budapest University of Technology and
  Economics, 1521 Budapest, Hungary}

\begin{abstract}
Dynamical phase transitions (DPT) are characterized by nonanalytical time evolution of the dynamical free energy. 
For general 2-band systems in one and two dimensions (eg. SSH model, Kitaev-chain, 
Haldane model, p+ip superconductor, etc.), we show that the time evolution of the dynamical free energy is crucially affected by the ground 
state topology of both 
the initial and final Hamiltonians, implying DPTs when the topology is changed under the quench. Similarly to edge states in topological 
insulators, DPTs can be classified as being topologically protected or not. In 1D systems the number of topologically protected 
non-equilibrium time scales are determined by the difference between the initial and final winding numbers, while in 2D no such relation 
exists for the Chern numbers. The singularities of dynamical free energy in the 2D case are qualitatively different from those of the 1D case, 
the cusps appear only in the first time derivative.
\end{abstract}
\date{\today}
\maketitle{}  
\bibliographystyle{apsrev}  


Topology\cite{hasankane,zhangrmp} and non-equilibrium dynamics\cite{polkovnikovrmp,dziarmagareview} are two vividly investigated fields of physics, with no strong bonds between them.
The  leading role played by topology in condensed matter has only been realized recently with the discovery of topological insulators, the descendants of quantum Hall states.
Some of their correlation functions are universal and are not influenced by the microscopic details of the system, but are rather determined by the underlying topology.
The analysis of non-equilibrium states, on the other hand, have emerged recently in a different field: in cold atomic systems. With the unprecedented control of preparing initial states and governing the time evolution,
a number of interesting phenomena has been observed such as the Kibble Zurek scaling \cite{LamporesiNatPhys2013}, the lack of thermalization in integrable systems \cite{KinoshitaNature2006}, etc.
 In this paper, we connect these two, seemingly unrelated fields and show that topology can be used as an organizing principle to classify out-of-equilibrium
systems.

The most popular setups for non-equilibrium dynamics are quench 
experiments in which the quantum system initially sits in the ground state of given Hamiltonian, but its time evolution is governed by another Hamiltonian. 
The quench protocol can  conveniently be characterized by the dynamical partition function with no reference to any particular observables, defined as 
\begin{equation} \label{eq:Zt}
Z(z)=\left\langle \psi \right| e^{- H z} \left|\psi \right\rangle \,,
\end{equation}
For positive real values of $z$ this gives the partition function of a field theory with boundaries $\left|\psi \right\rangle$ separated
by $z$ \cite{LeClairNuclPB1995}. For our purposes, we use  $z=it$ with $t$ real, which then gives the Loschmidt amplitude, that is, the  overlap of the time evolved state with the
initial state as $G(t)=Z(i t)$. It characterizes the time evolution and the stationary state after a long waiting time
\cite{Fagotti2013dynqpt}.
Similarly to the equilibrium situation, the dynamical free energy is defined as the logarithm per system
size $f(t)=-1/N^{d} \ln G(t)$.  In the thermodynamic limit it can be nonanalytical function of time, which was dubbed DPTs\cite{HeylPRL2013}. 
Although we are mostly interested in the dynamical partition function for imaginary arguments, following Fisher's method of studying phase transitions, its structures are revealed by analyzing the function on the whole complex plane. Fisher's method also supports
the analogy between phase transitions and DPTs. His idea was to study the zeros of the partition function, because
they are responsible for the possibly nonanalytic behaviour of the free energy \cite{Fisher1965}.

Former studies of DPTs mainly focused on spin systems  \cite{HeylPRL2013, Fagotti2013dynqpt, ObuchiPRE2012, SirkerPRB2014, KarraschPRB2013, 
HickeyPRB2013, HickeyarXiv2014XY, VajnaPRB2014XY, KrielDPT2014, CanoviDPT2014}. Here we extend the analysis of DPTs to general 2-band 
topological systems, including superconductors as well, where the topological number is the winding number or the Chern number, and demonstrate the prominent role of topology.

We consider 2-band translational invariant insulators and Bogoliubov-de Gennes superconductors in 1 and 2 dimensions. The Hamiltonian 
for these systems can be parametrized by a vector $\vd_{\vk}$,
\begin{align} \label{eq:Hamk}
H=\sum_{\vk} \vc_{\vk}^{+} h_{\vk} \vc_{\vk} &, & h_{\vk}=\vd_{\vk}\cdot\vsigma  \,,
\end{align}
where $\vc^{+}_{\vk}=(c_{\vk,A}^{+},\, c_{\vk,B}^{+})$ for insulators and $\vc^{+}_{\vk}=(c_{\vk}^{+},\, c_{-\vk})$ for superconductors. 
In the insulator case the internal degrees of freedom $A, B$ refer to pseudo-spin components, e.g. to different sublattices. 
A sudden quench protocol can be described by the change in the vector fields characterizing the Hamiltonian:
$\vd_{\vk}(t)=\vd^0_{\vk}$ for $t<0$ and $\vd_{\vk}(t)=\vd^1_{\vk}$ for $t>0$.
The Loschmidt amplitude following this quench can be expressed in a compact form independently of the spatial dimensions:
\begin{align} \label{eq:Gt_gen}
G(t)&=\prod_{\vk} \left[ \cos(\epsilon^1_{\vk} t)+ i \, \hvd^0_{\vk} \cdot \hvd^1_{\vk} \sin(\epsilon^1_{\vk} t)\right] \,,
\end{align}  
but the product is taken for all wavenumbers in the Brillouin zone for normal insulators and for half of the Brillouin zone 
for superconductors. Here,  $\hvd^i_{\vk}$ denotes the unit vector in the direction of  $\vd_{\vk}^i$ and $\epsilon^i_{\vk}=|\vd_{\vk}^i|$ for insulators 
and $\epsilon^i_{\vk}=2|\vd_{\vk}^i|$ for superconductors. 
The Fisher zeros, i.e. the solutions of $Z(z)=0$ are 
\begin{equation} \label{eq:zk_gen}
z_n(\vk)=\frac{i\pi}{\epsilon^1_{\vk}}(n+\frac{1}{2})-\frac{1}{\epsilon^1_{\vk}} \arth\left[\hvd^0_{\vk} \cdot \hvd^1_{\vk}\right],
\end{equation}  
which follow from the product form of the  Loschmidt amplitude (and hence the dynamical partition
function).
The Fisher zeros fill domains of the complex plane, which are indexed by an integer number $n$ and are parametrized with $\vk$. In 1 dimension these 
domains form lines, while in 2D they fill areas. The necessary condition to observe DPTs 
is having Fisher zeros approaching the imaginary axis, which occurs when $\vd^0_{\vk} \cdot \vd^1_{\vk} = 0$, i.e. when the $\vd_{\vk}$ vector 
in the final Hamiltonian is perpendicular to the initial one for some $\vk$. This geometrical condition connects DPTs with the topology 
of the initial and the final systems.
In the following we will consider the 1 and 2 dimensional cases separately.

\begin{figure}[h!]
\psfrag{d}[Bc][Bc][0.7]{$\,\vd_{\vk}^0$}
\psfrag{g}[Bc][Bc][0.7]{$\,\vd_{\vk}^1$}
\psfrag{a}[Bc][Bc][1.0]{(a)}
\psfrag{b}[Bc][Bc][1.0]{(b)}
\psfrag{c}[Bc][Bc][1.0]{(c)}
\centering
\includegraphics[width=2.8cm]{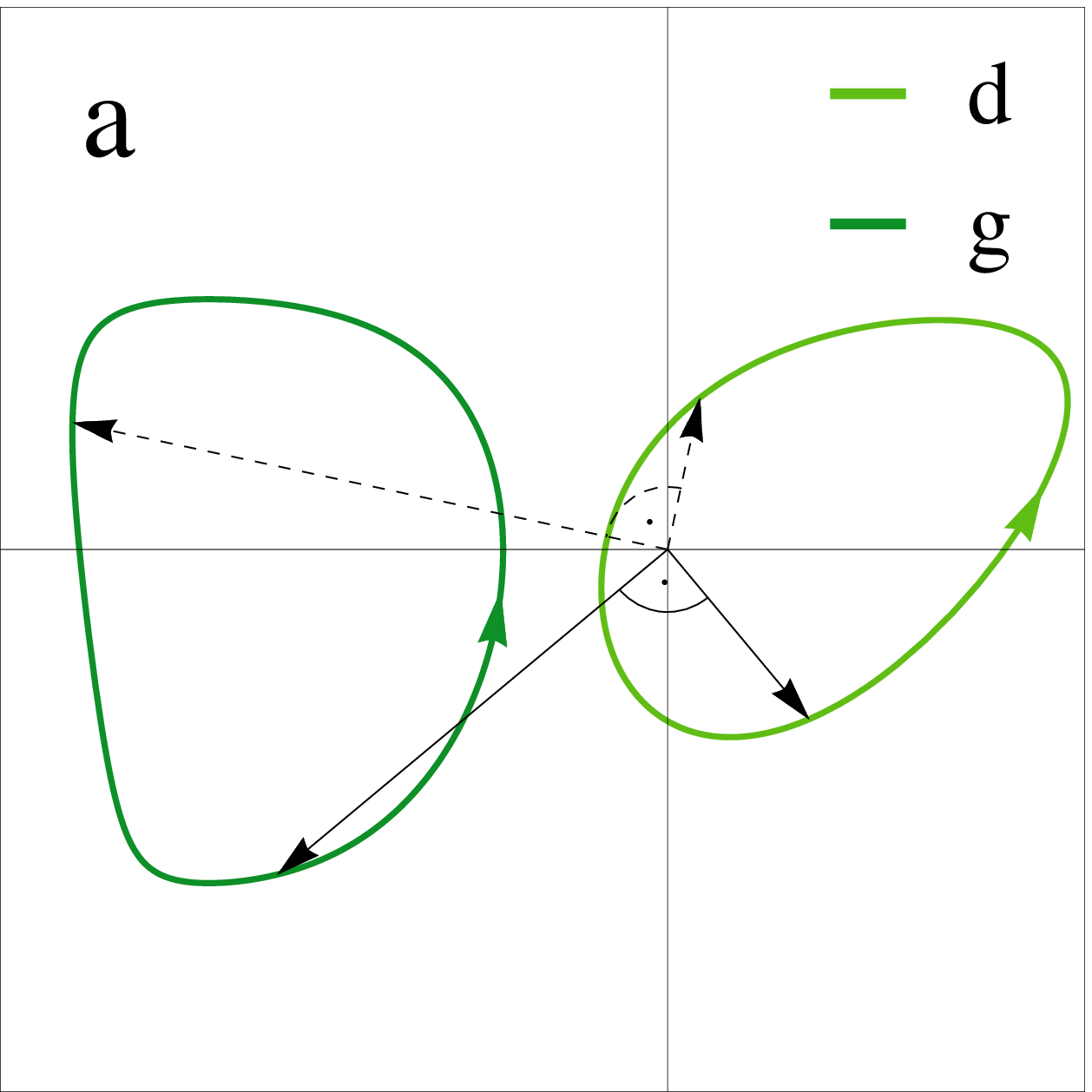}
\includegraphics[width=2.8cm]{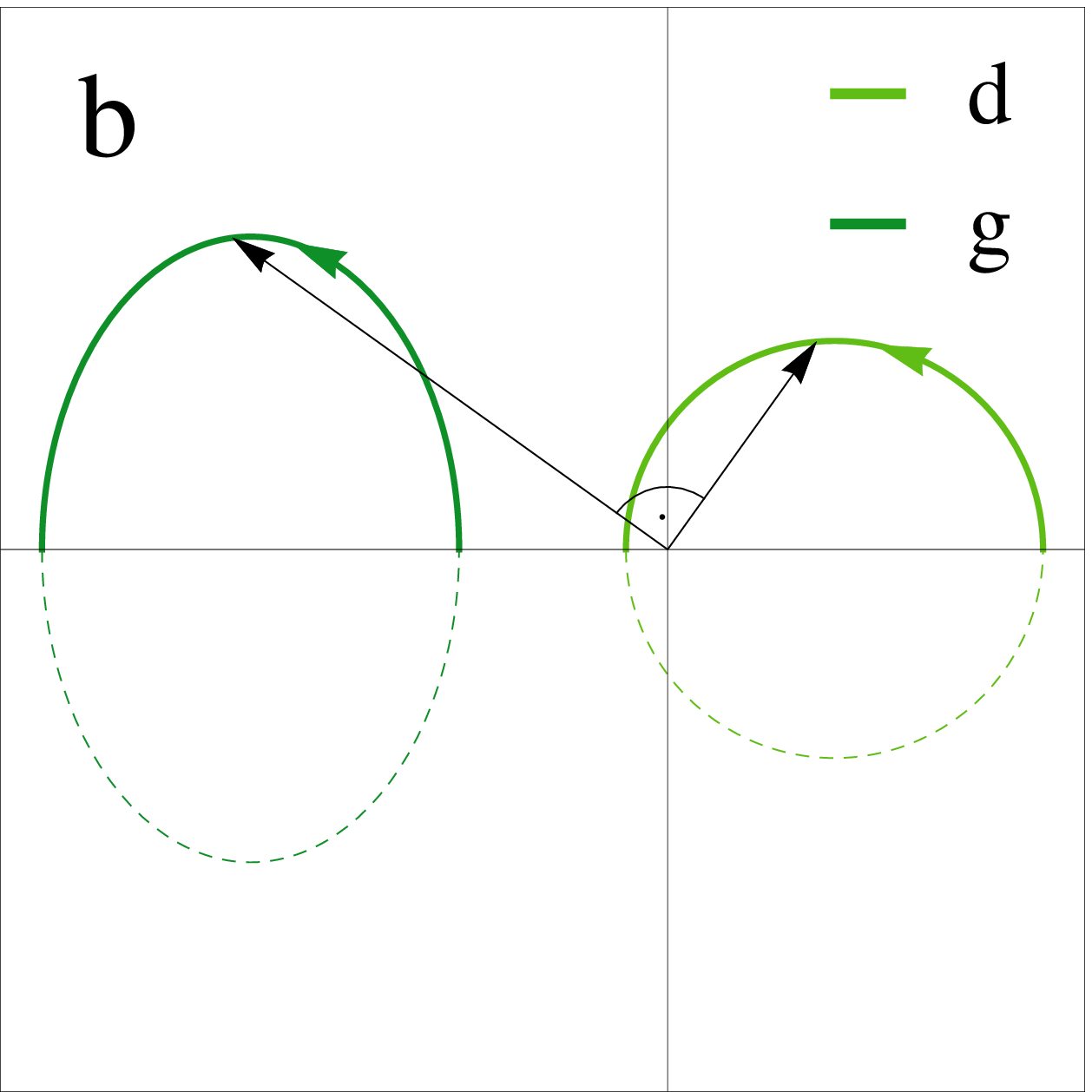}
\includegraphics[width=2.8cm]{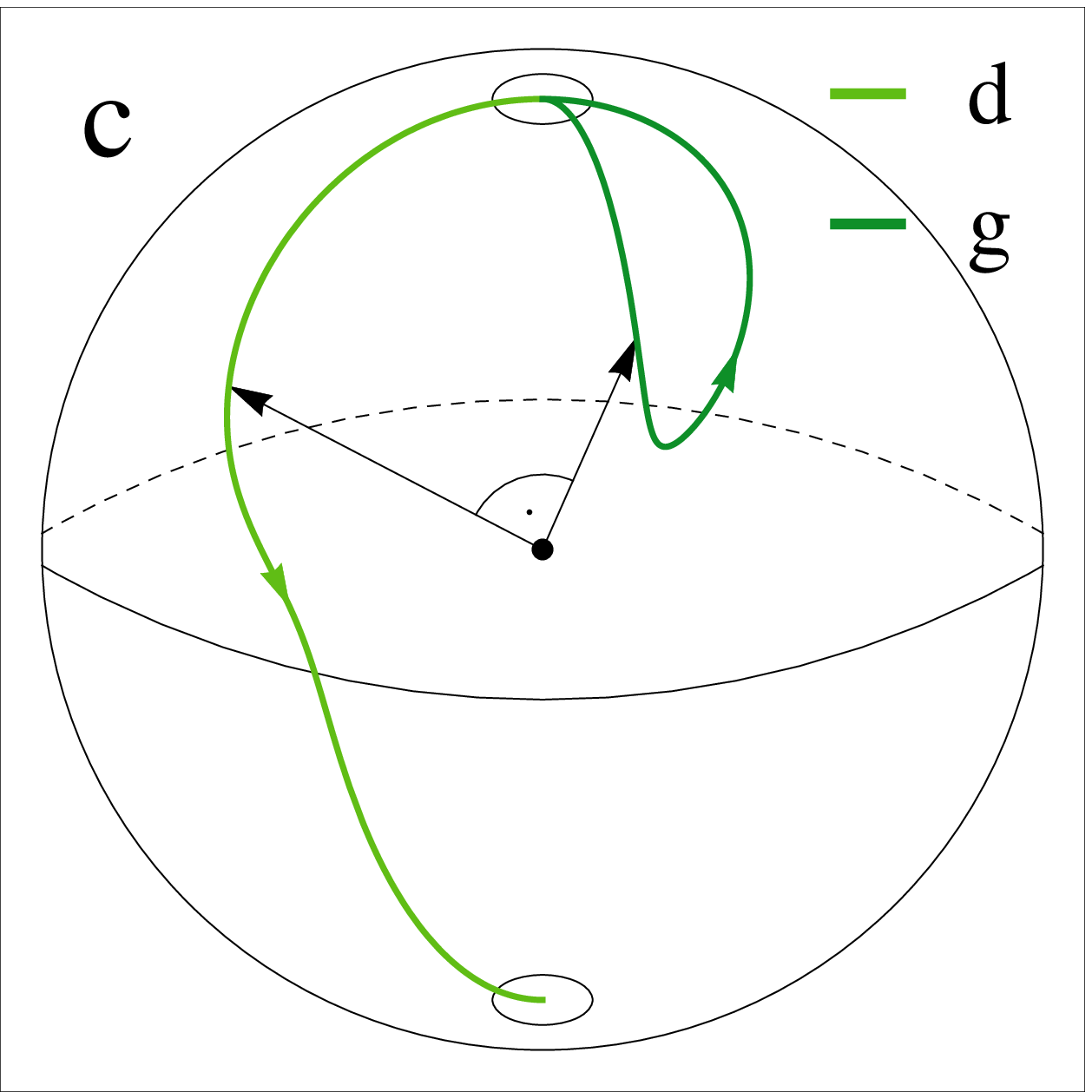}
\caption{Illustration for the existence of perpendicular vectors if the quench connects domains with different topological numbers.
(a) Topological insulators in AIII symmetry class. As $k$ goes through the Brillouin zone $\vd_k$ draws a closed loop. For any parametrization 
of these loops there will be at least $2 \Delta \nu$ wavenumbers for which  $\vd^0 \perp \vd^1$ if the winding number of the two vector fields 
differ by $\Delta \nu$.
(b) Superconductors in BDI symmetry class. In the $k\in[0,\pi]$ domain there will be at least $\Delta \nu$ perpendicular $\vd$ vectors in a quench 
characterized by $\Delta \nu$.
(c) Superconductors in D symmetry class. The vectors $\hvd$ are no longer confined to a plane, but the occurrence of perpendicular vectors is still 
ensured when the topology of $\hvd^1$ an $\hvd^0$ are different.}
\label{fig:windplot}
\end{figure}

\emph{One dimensional case}
Topological insulators in one dimension are characterized by chiral (AIII symmetry class) or chiral and particle-hole symmetry  
(BDI) \cite{RyuNJP2010}, which constrain $\vd_k$ to lie in a 2D plane. 
The corresponding topological number is the winding number: the number of times $\vd_k$ winds 
around the origin when $k$ sweeps through the Brillouin zone. If, for example $\vd_k$ lies in the $xy$ plane,
$\nu=\frac{1}{2 \pi}\int \mathrm{d}k \, (\hat{d}^x_{k}\partial_k \hat{d}^y_{k}-\hat{d}^y_{k}\partial_k \hat{d}^x_{k})$.
\begin{quote}
If the winding number of two vector fields $\vd^0_k$ and $\vd^1_k$ defined on the Brillouin zone ($S^1$) differ by $\Delta \nu \in \mathbb{N}$, 
 the image of the scalar product field $\hvd^0_{k} \cdot \hvd^1_{k}$ covers the interval $[-1,1]$ at least $2 \Delta \nu$ times.
\end{quote}
This means that the Fisher zeros in equation \eqref{eq:zk_gen} sweep through the real axis $2 \Delta \nu$ times while $k$ goes through 
the Brillouin zone.  Consequently there are at least $2 \Delta \nu$ points in $k$ space where the vectors are perpendicular 
(for illustration see FIG.~\ref{fig:windplot}), implying DPTs. 
Let us suppose that the (ground state) winding number of the initial (final) Hamiltonian is $\nu_0$ ($\nu_1$), then the angle of rotation 
$\phi^i_k$ for $\vd^i_k$ is a smooth function differing by $2\pi \nu_i$ at $k=-\pi$ and $\pi$ for $i=0,1$. The angle of rotation $\Delta\phi_k$ 
between $\vd^0_k$ and $\vd^1_k$ changes $2\pi \Delta \nu$, hence $\hvd^0_{k} \cdot \hvd^1_{k}= \cos(\Delta\phi_k)$ 
covers the interval $[-1,1]$ at least $2 \Delta \nu$ times. 
If the model has further symmetries that connect wavenumbers $k$ and $-k$ (e.g. inversion or time reversal symmetry (TRS)), the Fisher lines 
can be doubly degenerate and there will be only $\Delta \nu$ non equilibrium time scales, as it happens for example in the SSH model (see later).

Our argument applies also for 1D topological superconductors (e.g. the Kitaev chain and its generalization for higher winding numbers 
\cite{NiuPRB2010}) with a little modification. In that case the product is taken only for positive momenta in the Loschmidt 
amplitude in Eq.~\eqref{eq:Gt_gen}. The Bogoliubov Hamiltonian is particle-hole symmetric (PHS) by construction, implying that the $x$ and $y$ components of $\vd_k$ are odd 
and the $z$ component is even function of the wavenumber:
$d^x_{-k}=-d^x_k$, $d^y_{-k}=-d^y_{-k}$ and $d^z_{-k}=d^z_{k}$.
For the degenerate momenta 
$k=0,\pi$, the vector describing the Hamiltonian points to the $z$ direction:
$\vd_{0/\pi}=(0,0,d^z_{0/\pi})$.
If the system is time reversal invariant as well (BDI symmetry class)
$d^x_k \equiv 0$ 
and the topological number is the winding number 
similarly to the previous case. Because of PHS the winding of the angle of $\vd_k$ is already determined in the $k\in(0,\pi)$ domain. That is, 
if the winding number is $\nu$, the angle changes $\pi \nu$ while $k$ goes through the positive half of the Brillouin zone. Therefore for a quench 
from $\nu_0$ to $\nu_1$ DPTs will appear with $\Delta \nu$ topologically protected time scales. This result applies for the  previous studies in 
the literature for the transverse field Ising model and for the quantum XY chain \cite{HeylPRL2013, VajnaPRB2014XY}, which can be mapped 
to Hamiltonians in the form of \eqref{eq:Hamk}.

If TRS is broken (D symmetry class, \cite{RyuNJP2010}) $\vd_k$ is not confined to a 2D plane. The $\mathbb{Z}_2$ invariant is $0$ 
(topologically trivial) if 
$\hvd_{0}=\hvd_{\pi}(=(0,0,\pm 1))$ and it is $1$ (nontrivial) if $\hvd_{0}=-\hvd_{\pi}$. If the quench connects phases with different 
topology, e.g. $\nu_0=1$ and $\nu_1=0$, there must be a wavenumber $k^{*}$ for which $\vd_{k^{*}}^0 \cdot \vd_{k^{*}}^1=0$ , 
because $\hvd_0^0 \cdot \hvd_0^1=s$ and $\hvd_{\pi}^0 \cdot \hvd_{\pi}^1=-s$ with $s=\pm 1$, 
hence $\hvd_{\vk}^0 \cdot \hvd_{\vk}^1$ covers the interval $[-1,1]$.

So far we have demonstrated that the change in topology under a sudden quench  is an eligible condition for DPTs to occur.  
Note that $\vd^0$ and $\vd^1$ can become perpendicular \emph{accidentally} even if the topological numbers do not differ in the initial 
and final Hamiltonian \cite{VajnaPRB2014XY}, but in that case there is no guarantee that it will happen. 
This parallels to the appearance of topologically non-protected edge or surface states in certain systems, 
whose existence is not connected to topology but is accidental \cite{hasankane,RyuPRL2002,KuramotoPRB2011}.

\emph{Two dimensions}
The topological number is the Chern number $Q$ for 2-band topological insulators, which is calculated from the vector $\vd_{\vk}$ defining the Hamiltonian \cite{ZhangPRB2006},
\begin{align} \label{eq:Chern}
Q=\frac{1}{4\pi}\int_{BZ} \mathrm{d}k_x \mathrm{d}k_y \,\, \hvd_{\vk} \cdot ( \partial_{k_x} \hvd_{\vk} \times \partial_{k_y}\hvd_{\vk} )
\end{align}
counting how many times the surface defined by $\hvd_{\vk}$ covers the unit sphere.
We show that if the quench connects 
phases with Chern numbers differing in their moduli, DPTs will necessarily occur. However, DPTs in 2 dimensions are qualitatively 
different from those in 1D, because of the Fisher zeros form areas instead of lines. 

\begin{quote}
If the Chern numbers of two vector fields $\vd^0_{\vk}$ and $\vd^1_{\vk}$ defined on the Brillouin zone ($T^2$) differ in the modulus 
$|Q_{1}|\neq |Q_{0}|$, the image of the scalar product field $\hvd^0 \cdot \hvd^1$ is $[-1,1]$.
\end{quote}
We prove our statement in four steps:\\
(a) If $\hvd^0_{\vk} \cdot \hvd^1_{\vk}>-1$ $\Rightarrow$ $Q_1=Q_0$, because there is a continuous mapping $\vec{f}_{\vk}(\gamma)$ between 
$\hvd^0_{\vk}$ and $\hvd^1_{\vk}$ such that $|\vec{f}_{\vk}(\gamma)|>0$. 
\begin{align*}
\vec{f}_{\vk}(\gamma)&=(1-\gamma) \hvd^0_{\vk} + \gamma  \hvd^1_{\vk} , \qquad \gamma \in [0,1]\\
|\vec{f}_{\vk}(\gamma)|^2&=1+2\gamma(1-\gamma) (-1+ \hvd^0_{\vk} \cdot \hvd^1_{\vk}) {>0}
\end{align*}
The inequality in the second line came from the fact that $\gamma(1-\gamma)<1/4$ for $\gamma \in [0,1]$.
In other words if the vector fields $\hvd^0_{\vk}$ and $\hvd^1_{\vk}$ are nowhere antiparallel, then one can continuously deform one into the other. Under 
this deformation the Chern number does not change \cite{Nakahara2003}. \\
(b) If  $Q_1\neq Q_0$ $\Rightarrow$ $\hvd^0_{\vk} \cdot \hvd^1_{\vk}=-1$ for some $\vk$.
This comes from reversing (a) and can be proved indirecly.\\
(c) If $Q_1\neq -Q_0$ $\Rightarrow$ $\hvd^0_{\vk} \cdot \hvd^1_{\vk}=1$ for some (other) $\vk$.
We trace back this statement to (b) by defining the vector field $\hat{\vd}'^{1}_{\vk}\equiv -\hvd_{\vk}^{1}$, 
which satisfy $Q'_1 \neq Q_0$.  Hence $\hvd^0_k \cdot \hat{\vd}'^{1}_{\vk}=-1$ for some $\vk$ implying (c). 
Supported by the continuity of $\hvd^i_{\vk}$ when the topology is well-defined, combining (b) and (c) finishes the proof.
 
This argument cannot be generalized to the $Q_1=-Q_0$ case, a trivial counterexample is given by the quench $\vd^1_{\vk}=-\vd^0_{\vk}$, 
where the initial 
and final Chern numbers are the opposite, but $\vd^0 \cdot \vd^1 \equiv -1$. 

The statement ensures that Fisher zeros connect $-\infty$ to $\infty$ if the modulus of the Chern number changes under the quench. 
Nevertheless, one might find Fisher lines connecting $-\infty$ to $\infty$ also when the modulus of the Chern numbers are the same. 
A 2D system can be thought of as a collection of 1D chains. If these 1D systems can be characterized by winding numbers, it is enough to 
find a pair of these 1D systems with differing winding numbers to see DPTs. On the other hand, Fisher zeros can also expand through the 
imaginary axis \emph{accidentally} similarly to the 1D case.    

In the superconducting case the product in Eq.~\eqref{eq:Gt_gen} is taken for the half Brillouin zone. However, because of PHS
one gets exactly the same contribution from the other half of Brillouin zone, so one can express $G(t)^2$ as a product over the whole BZ. 
From this the existence of DPTs follows for quenches connecting superconducting phases with different moduli of the Chern numbers. 

Having established our main results, a few examples follow. 

\begin{figure}[h!]
\psfrag{r}[Bc][Bc][0.8]{$\Re\{z\}$}
\psfrag{i}[Bc][Bc][0.8]{$\Im\{z\}$}
\psfrag{f}[Bc][Bc][0.8]{$\Re\{f(t)\}$}
\psfrag{a}[Bc][Bc][1.0]{(a)}
\psfrag{b}[Bc][Bc][1.0]{(b)}
\psfrag{t}[Bc][Bc][1.0]{t}
\psfrag{-}[Br][Bc][0.8]{$-$}
\centering
\includegraphics[height=4cm]{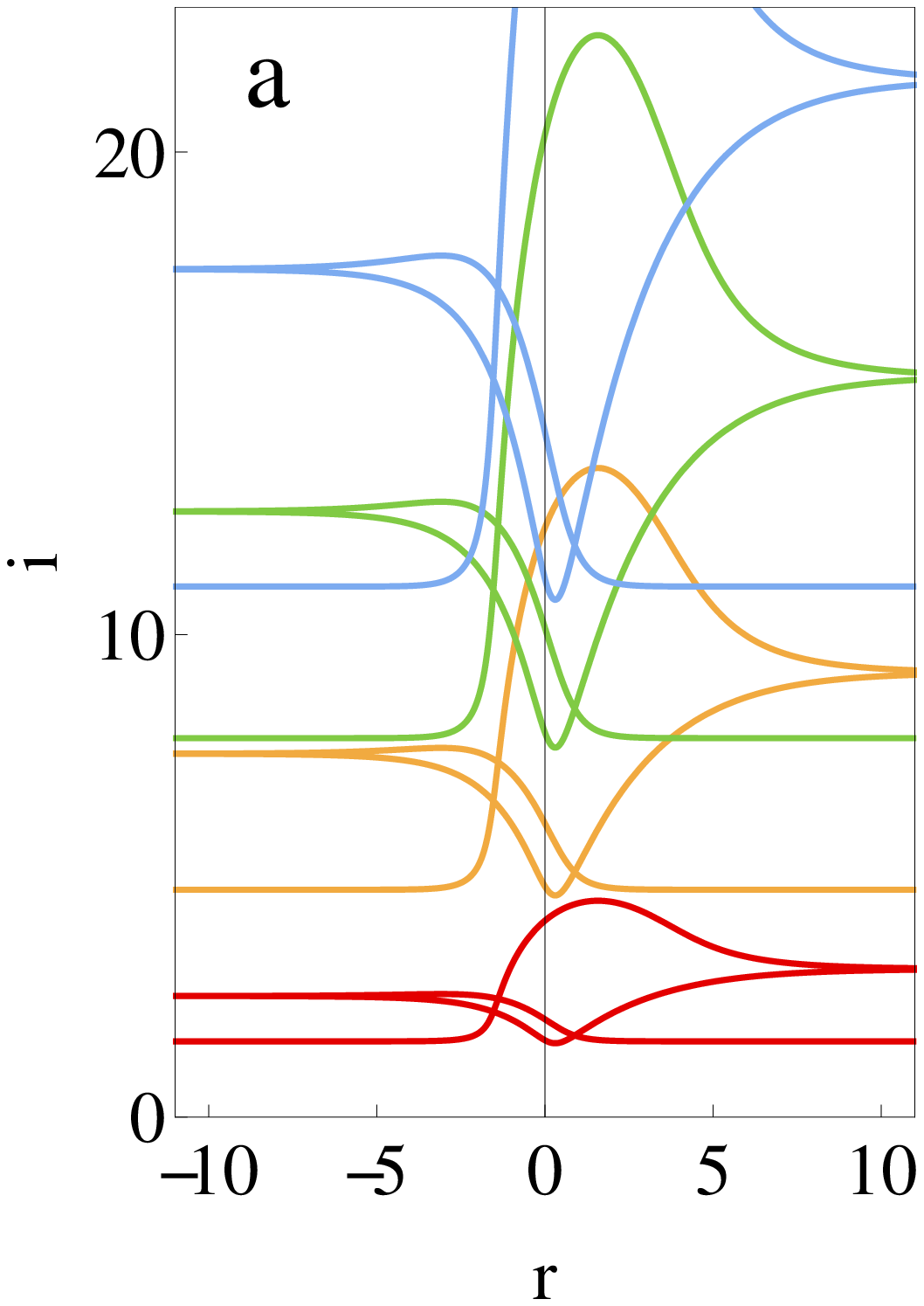}
\includegraphics[height=4cm]{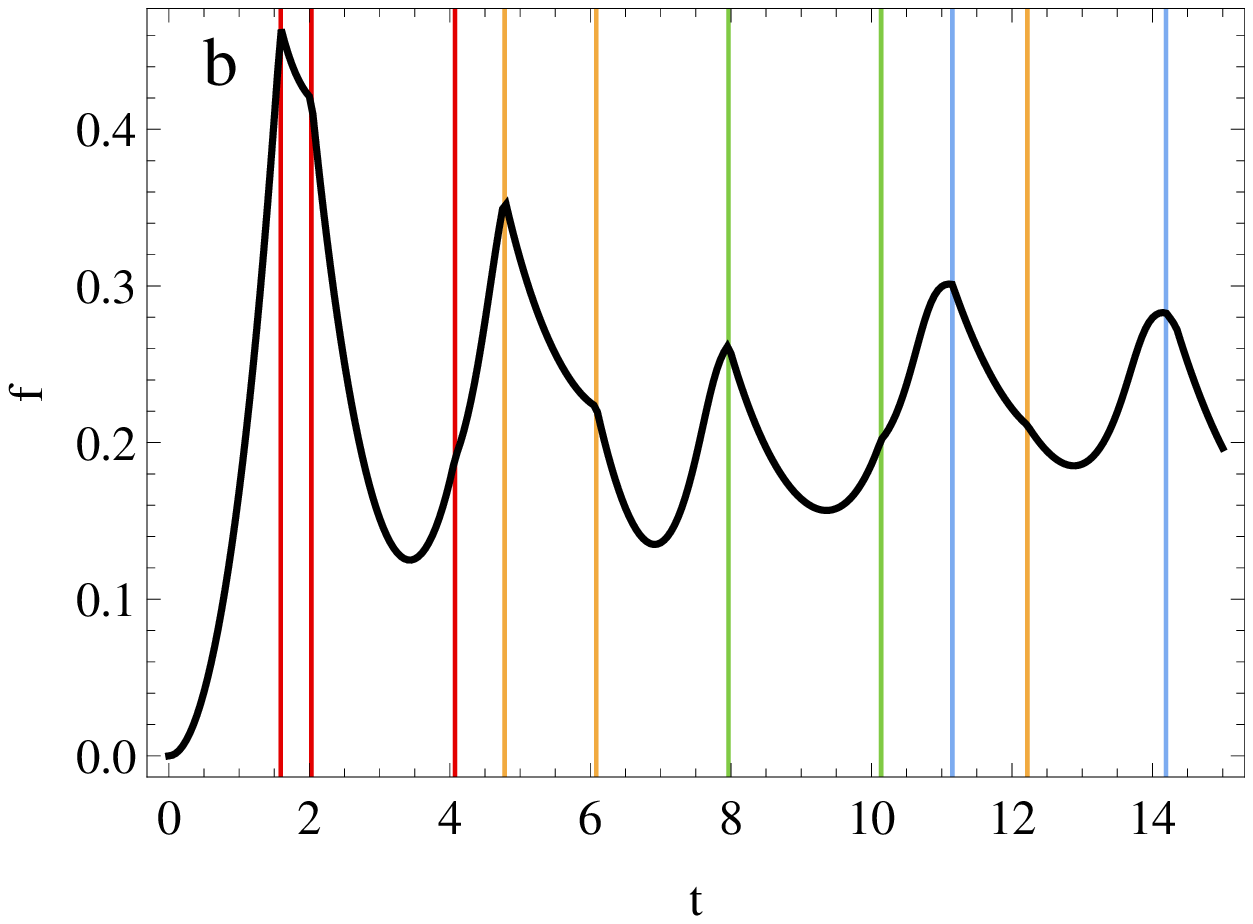}
\caption{Fisher lines $z_n(k)$ (for $n=0..3$) and DPTs in the generalized SSH model in a quench $\nu_0=1$ $\rightarrow$ $\nu_1=-2$. 
(a) The lines of Fisher zeros are doubly degenerate due to the $(k, -k)$ symmetry, they sweep through the real axis $|\Delta\nu|=3$ times. 
(b) Dynamical phase transitions appear where the Fisher zeros cross the imaginary axis. The grid lines show the DPTs corresponding to the 
first four Fisher lines.}
\label{fig:SSHplot}
\end{figure}

\emph{Generalized SSH model}
The SSH model is a 1D tight-binding chain that was originally introduced to model polyacetylene \cite{SSHPRL1979}. It is probably the simplest
topological insulator, belonging to the BDI symmetry class \cite{RyuNJP2010}. The model is described by 
$\vd_k=(t_{0}+t_{-1} \cos(k),t_{-1} \sin(k),0)$, where $t_0$ and $t_{-1}$ are the staggered hopping amplitudes. The ground state is topologically 
trivial ($\nu=0$) when $t_0>t_{-1}$ and is non-trivial if $t_0<t_{-1}$. The model can be extended to produce higher winding numbers by introducing 
longer ranged hopping terms that preserve chiral symmetry. The Hamiltonian in this case is characterized by the vector 
\begin{equation}
\vd_k=(t_0'+\sum_{m=1}^{\infty} t_m' \cos(m k),\sum_{m=1}^{\infty} t_m'' \sin(m k),0) \,,
\end{equation}
where $t_0'=t_0$, $t_m'=t_m+t_{-m}$, $t_m''=t_{-m}-t_m$, and $t_m$ is the real hopping amplitude between sublattices 
$A$ in unit cell $i$ and $B$ in unit cell $i+m$. We note that in this labeling of 
the hopping amplitudes $t_m$ and $t_{-m}$ are independent and are responsible for the staggered nature of the system. 
Higher winding numbers can be produced by the proper choice of the hopping amplitudes, for example the winding number is $|\nu|=n\geq 1$ 
if $t_n'$ and $t_n''$ dominate the other hopping terms. An eligible condition for this is 
$|t_0'|+\sum_{m=1}^{m \neq n}{|t_m'|+|t_m''|}< \min\{|t_n'|,|t_n''|\}$. 
Besides the chiral symmetry this model has TRS as well. Therefore in a quench characterized by $\Delta \nu$, the 
$2 \Delta \nu$ topologically protected $(-\infty,\infty)$ sections of the Fisher lines are pairwise degenerate 
(consider e.g. that $|\vd^1_k|=|\vd^1_{-k}|$ and $\vd^0_k \cdot \vd^1_k=\vd^0_{-k} \cdot \vd^1_{-k}$), implying only $\Delta \nu$ non-equilibrium 
time scales.  The flow of the Fisher lines and the dynamical free energy are shown on FIG.~\ref{fig:SSHplot} for a quench from a phase with 
$\nu_0=1$ to $\nu_1=-2$. 

\emph{The Haldane model} is a next-nearest neighbor hopping model on a honeycomb lattice with artificial magnetic field \cite{HaldanePRL1988}, which can produce topologically nontrivial states. It is characterized by 
\begin{equation}
\vd_{\vk}=(\Re\{f(k)\},\Im\{f(k)\},m-g^{\text{asym}}(k,\phi))\,
\end{equation} 
where $f(k)=\gamma_1 \sum_j e^{-i \vk \vdelta_j}$ and vectors $\vdelta_j$ point to the three nearest neighbors. The mass term $m \sigma_z$ describes 
a homogeneous staggered lattice potential. The $-g^{\text{asym}}(\vk,\phi) \sigma_z$ term comes from a second neighbor hopping considering the 
staggered magnetic field characterizing the Haldane model. This latter term is necessary to produce nontrivial topology in the model.
The Chern number depends on the phase $\phi$ characterizing the magnetic field, on the next nearest hopping amplitude $\gamma_2$, and on the mass 
term. The Chern number is $Q=0$ if $|m|>|3\sqrt{3}\gamma_2 \sin{\phi}|$, and $Q=\pm 1$ if $|m|<|3\sqrt{3}\gamma_2 \sin{\phi}|$ with the sign 
depending on $\phi$ and $\gamma_2$. 

\begin{figure}[h!]
\centering
\psfrag{r}[Bc][Bc][1.0]{$\Re\{z\}$}
\psfrag{i}[Bc][Bc][1.0]{$\Im\{z\}$}
\psfrag{j}[Bc][Bc][0.8]{$\Im\{z\}$}
\psfrag{f}[Bc][Bc][1.0]{$\Re\{f'(t)\}$}
\psfrag{a}[Bc][Bc][1.0]{(a)}
\psfrag{b}[Bc][Bc][1.0]{(b)}
\psfrag{t}[Bc][Bc][1.0]{t}
\psfrag{w}[Bc][Bc][0.8]{$\Re\{z\}=0$}
\psfrag{d}[Bc][Bc][0.8][90]{$\rho(0,\Im\{z\})$}
\psfrag{-}[Br][Bc][1.0]{$-$}
\includegraphics[width=8.5cm]{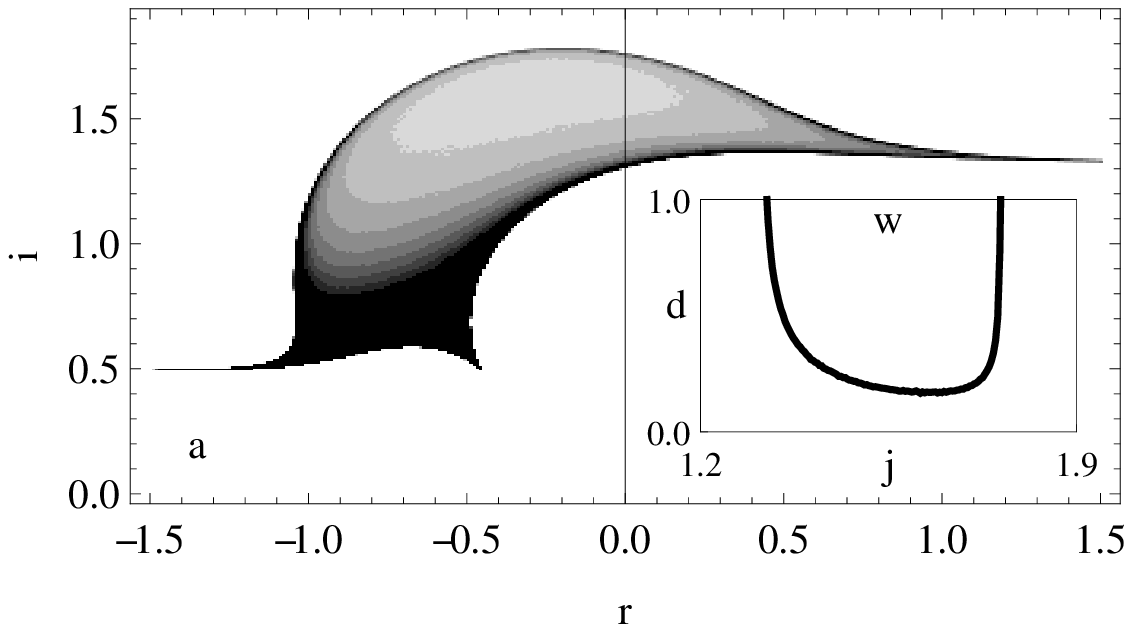}\\
\includegraphics[width=8.0cm]{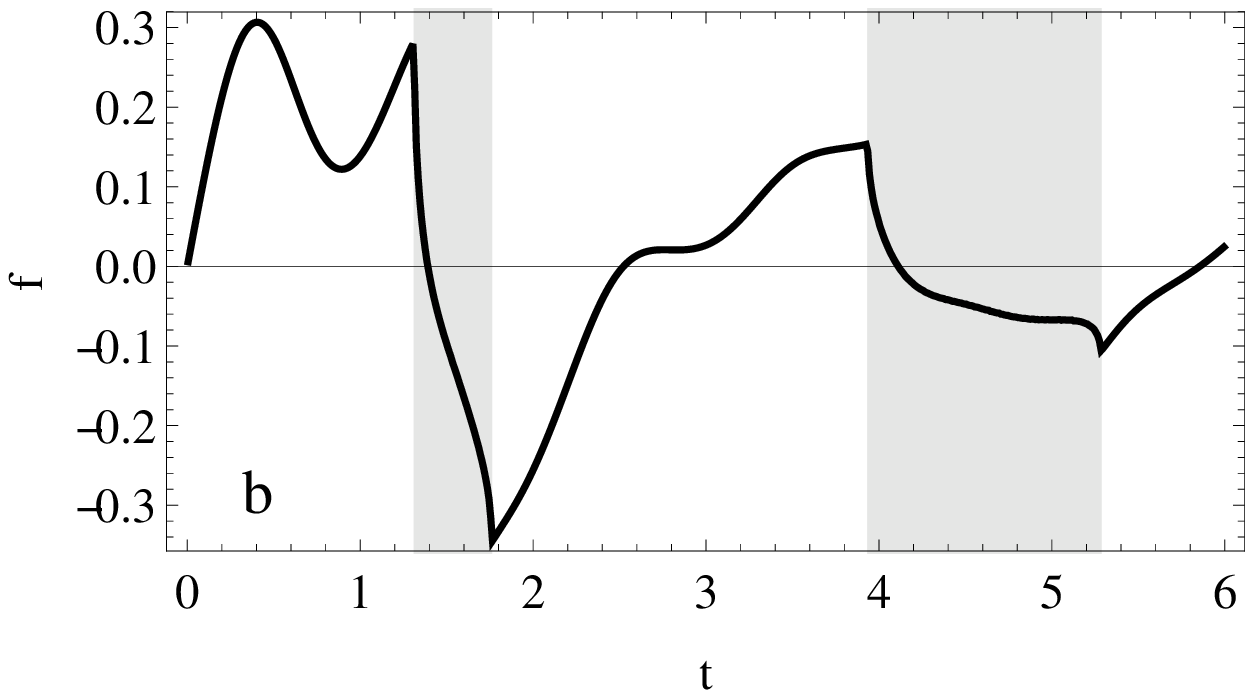}
\caption{Quench in the Haldane model from $Q_0=0$ to $Q_1=1$. 
(a) Fisher zeros corresponding to $n=0$ in Eq.~\eqref{eq:zk_gen}. 
The normalized density of the Fisher zeros $\rho(\Re\{z\},\Im\{z\})$ is shown as the darkness of the area. 
Inset: the density diverges on the imaginary axis at the boundary of the Fisher area.   
(b) Cusplike singularities in the first derivative of the dynamical free energy. 
The shaded areas emphasize the regions where the Fisher zeros cross the time axis.}
\label{fig:haldaneFD}
\end{figure}

We have already proved that if the Chern number of the initial and final Hamiltonian differs, the Fisher zeros connect $-\infty$ with $\infty$. 
However, in contrast to the one dimensional case, in two dimensions the Fisher zeros fill areas rather than forming lines.   
Similar behaviour occurs for  quenches in spin-glass systems \cite{ObuchiPRE2012,ObuchiJP2013}. In our case the appearance of 
Fisher area is not unexpected, since each Fisher domain corresponding to a given $n$ in Eq.~\eqref{eq:zk_gen} is parametrized by two variables 
$k_x$ and $k_y$. In contrast to Fisher lines if a Fisher area crosses the imaginary axis, the dynamical free energy looks smooth and its first 
derivative shows cusps at the boundaries of the Fisher area (FIG.~\ref{fig:haldaneFD}(b)). This is  understood by expressing the 
singular part of the dynamical free energy with the Fisher zeros \cite{HeylPRL2013}
\begin{align} \label{eq:fz_fisher}
f^s(t)=-\lim_{N\rightarrow \infty}\frac{1}{N^{d}}\sum_{n,k} \ln\left(1-\frac{i t}{z_{n}(k)}\right).
\end{align} 
If an area with homogeneous density of Fisher zeros cross the imaginary axis, the second derivative of the dynamical 
free energy will jump at the boundary of the Fisher-area. The size of the jump is proportional to the density of zeros 
normalized by the system size ($\rho$) and with the cosine square of the impact angle of the boundary line ($\phi$). 
\begin{align} 
\lim_{\epsilon \rightarrow 0+} f''(t_0+\epsilon)-f''(t_0-\epsilon)=-2 \pi D \cos^2 \phi
\end{align}
If the density of the Fisher zeros diverge as $|y-y_0|^{-\alpha}$ at the boundary of the Fisher area, then the slopes of the cusps in $\Re\{f'(t)\}$ 
inside the Fisher area diverge similarly. In the Haldane model this latter behavior occurs as shown in FIG.~\ref{fig:haldaneFD}. 

These  features are not specific to the Haldane model, but show up in a wide range of models, e.g.  in the "half"-BHZ model \cite{BHZSci2006}, 
which is described by $\vd_{\vk}=(A \sin{kx},A \sin{ky}, \Delta+\cos{k_x}+ \cos{k_y})$, 
or in the lattice version of the chiral topological $p+ip$ superconductor with similar $\vd_{\vk}$ \cite{BuhlerNCOMM2014}.

\emph{Conclusion.} We have demonstrated that if the topological number changes under the quench, 
DPT has to occur. Previously the existence of DPTs was analyzed in comparison with equilibrium phase 
transitions \cite{HeylPRL2013,KarraschPRB2013,SirkerPRB2014,VajnaPRB2014XY}. 
Based on our results the topological nature of the equilibrium phase transition should be investigated as well. 
Our preliminary numerical simulations indicate that a change in the 
topological number under the quench protocol implies DPTs in the disordered case as well, which deserves further attention together with 
 systems with more (than 2) bands. 
DPTs in two dimensional systems are qualitatively different from those in one dimension as the cusps in the dynamical 
free energy appear in the first time-derivative. The distinct behaviour of the Fisher zeros together with different
types of nonanalyticities in various systems may open a path to define universality classes in DPTs.  

\begin{acknowledgments}

This research has been  supported by the Hungarian Scientific  Research Funds Nos. K101244, K105149, K108676, by the ERC Grant Nr. ERC-259374-Sylo and 
by the Bolyai Program of the Hungarian Academy of Sciences.
\end{acknowledgments}

\bibliographystyle{apsrev}

\bibliography{refgraph}

\end{document}